# Hot exciton transport in WSe$_2$ monolayers

Darwin F. Cordovilla Leon[1,†], Zidong Li[2,†], Sung Woon Jang[2] and Parag B. Deotare[1, 2,‡]

[1]Applied Physics Program, University of Michigan-Ann Arbor, 450 Church St., 1425 Randall Laboratory, Ann Arbor, MI 48109, USA

[2]Electrical and Computer Engineering Department, University of Michigan-Ann Arbor, 1301 Beal Ave., Ann Arbor, MI 48109, USA

We experimentally demonstrate *hot* exciton transport in h-BN encapsulated WSe$_2$ monolayers via spatially and temporally resolved photoluminescence measurements at room temperature. We show that the nonlinear evolution of the mean squared displacement of the non-resonantly excited *hot* exciton gas is primarily due to the relaxation of its excess kinetic energy and is characterized by a density-dependent fast expansion that converges to a slower, constant rate expansion. We also observe saturation of the *hot* exciton gas' expansion rate at high excitation densities due to the balance between Auger-assisted *hot* exciton generation and the phonon-assisted *hot* exciton relaxation processes.

Exciton transport in transition metal dichalcogenide (TMD) monolayers has received significant attention[1–9] due to the prospects of TMD-based excitonic devices operating at room-temperature[6,10]. While several of these studies have focused on quantifying the transport properties of excitons in TMD monolayers[1,3,4,7,8], the origin of the apparent time-dependent exciton diffusivity observed in these materials[7,8] still remains unexplained.

Time-varying diffusivities are typically associated with anomalous diffusive transport of a distribution of particles, which is characterized by a nonlinear evolution of the distribution's mean squared displacement (MSD)[11–14]. While numerous processes in nature appear to follow anomalous diffusive behavior[15–21], the physical mechanisms that lead to such phenomena are unique to each particular system.

In layered semiconductors, the reduced Coulomb screening enhances many-body interactions, which can radically influence the dynamics of excitons, especially in the high excitation density regime[22–26]. In this work, we show that the relaxation of the kinetic energy of a *hot* gas of excitons formed by a non-resonant optical excitation and exciton-exciton Auger scattering are responsible for the nonlinear evolution of the MSD of excitons in h-BN encapsulated WSe$_2$ monolayers. Specifically, we explore the effect of excitation density and photon energy on the temperature of the exciton gas by monitoring the temporal evolution of the photoluminescence's (PL) spatial profile of the exciton gas at room temperature. Our results demonstrate that the initial fast expansion of the exciton gas is the result of *hot* exciton transport and has a minimal contribution from Auger broadening. Auger broadening refers to the apparent fast diffusion that results from the center of the exciton PL intensity profile dropping faster than the edges due to density dependent non-radiative Auger recombination[3].

WSe$_2$ monolayers were mechanically exfoliated from a bulk WSe$_2$ crystal and encapsulated with hexagonal boron nitride (h-BN). More details of the fabrication process can be found in the supplementary information. The PL emission of excitons in h-BN encapsulated WSe$_2$ monolayers was monitored using the technique described in references[7,27,28], where a temporally and spatially resolved map of the WSe$_2$ monolayers' PL was constructed following a pulsed laser excitation with a Gaussian intensity profile. These temporally-resolved spatial maps were built one pixel at the time by using time-correlated single photon counting while scanning an avalanche photodiode detector across the PL emission spot.

Typical optical measurements on TMD monolayers employ excitations with photon energies much higher than these materials' bandgaps. This type of excitation creates electron-hole pairs with excess energy that can be relaxed via ultrafast interaction with phonons and eventually form excitons with high center-of-mass kinetic energy and momentum[29]. Excitons with high kinetic energy have a high probability of non-radiative recombination which takes place either due to exciton dissociation, trapping by defects[29–31] or rapid exciton-exciton Auger scattering[3,24,25,31–34]. Auger scattering is much more likely to occur in systems with low dimensionality, such as WSe$_2$ monolayers, where the reduced Coulomb screening enhances many-body interactions[3,24,25,31–35]. Depending on the relative recombination rates, a fraction of the initially created *hot* gas completes the kinetic energy relaxation during its lifetime via exciton-phonon scattering processes[29,36]. Eventually, the relaxed or *cold* excitons recombine radiatively.

---

† These authors contributed equally

‡ Corresponding author: pdeotare@umich.edu

In a system of *cold* excitons moving in an energetically homogeneous medium, the MSD of a Gaussian distribution of such excitons, denoted by $\langle \Delta\sigma^2(t)\rangle$, evolves linearly with time according to $\langle \Delta\sigma^2(t)\rangle = 2Dt$, where $D$ is the diffusion coefficient or diffusivity[37]. However, if the evolution of the MSD is nonlinear, it is typically described by the power law model $\langle \Delta\sigma^2(t)\rangle = \Gamma t^\alpha$, where $\Gamma$ and $\alpha$ are known as the transport factor and anomalous coefficient respectively[12,14,18,38–41]. Any deviation from $\alpha = 1$ is known as anomalous diffusion[38]. Anomalous-diffusive motion of carriers and excitons has traditionally been associated with hopping transport between localized states in solids with high energetic disorder[15–19,41,42]. This type of transport is negligible in h-BN encapsulated TMD monolayers as it has been shown that encapsulation reduces energetic disorder as well as surface roughness scattering, charged impurity scattering, and exciton-exciton scattering[43–45].

Figure 1 shows the time evolution of the MSD of excitons in an h-BN encapsulated WSe$_2$ monolayer for increasing excitation densities. We observe a progression of the MSD evolution from linear to nonlinear as the excitation density increases. The nonlinear regime is characterized by a rapid rise of the MSD at early times, followed by a transition into a slower, constant rate expansion for timescales beyond hundreds of picoseconds. The slope the MSD at early times increases with excitation density and saturates at elevated densities. The initial exciton diffusivities range from values between 0.5 cm$^2$s$^{-1}$ at low excitation and 4 cm$^2$s$^{-1}$ at high excitation densities (see supplementary Figure S4). We attribute the apparent anomalous diffusive motion of excitons in h-BN encapsulated WSe$_2$ monolayers to the kinetic energy relaxation of *hot* excitons formed by non-resonant optical excitation[46–55] and Auger broadening.

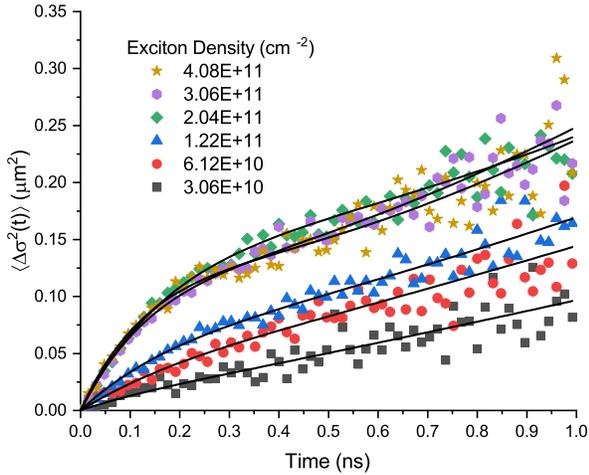

Figure 1: Evolution of the MSD of excitons in a h-BN encapsulated WSe$_2$ monolayer on a SiO$_2$/Si substrate for increasing excitation densities. The MSD is defined as $\langle \Delta\sigma^2(t)\rangle \equiv \sigma^2(t) - \sigma^2(0)$ where $\sigma(t)$ represents the standard deviation of the Gaussian distribution of excitons at time $t$. The solid lines represent the kinetic energy relaxation model fit represented by Eq.(3) and the markers are the experimental data. The results of these fits are shown in Figure S3 of the supplementary information.

A fraction of the fast rise in MSD can be attributed to Auger broadening, which can be significant at high excitation densities[3,50,56]. In this process, the PL intensity at the center of the emission spot, where the exciton density is the highest, decreases much more rapidly than the edges due to nonradiative exciton-exciton Auger recombination[3]. Such "flattening" of the PL emission profile can be misinterpreted as spatial exciton diffusion. To quantify the contribution of Auger broadening, we analyzed the time resolved PL (TRPL) as shown in Figure 2(a). Specifically, the evolution of the TRPL was modeled using the rate equation

$$\frac{\partial}{\partial t}n(t) = -\frac{n(t)}{\tau} - \gamma_A n^2(t) \quad (1)$$

where $n(t)$, $\tau$ and $\gamma_A$ represent the exciton density, lifetime and Auger constant respectively[32]. The data in Figure 2(a) was fitted (solid lines) with the solution to this rate equation given by

$$n(t) = \frac{n(0)e^{-t/\tau}}{1 + [n(0)\gamma_A \tau](1 - e^{-t/\tau})} \quad (2)$$

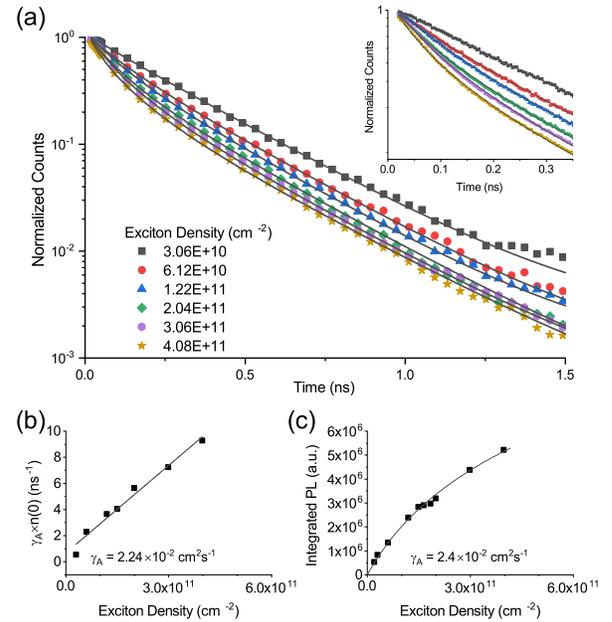

Figure 2: (a) Time-resolved photoluminescence (TRPL) of a h-BN encapsulated WSe$_2$ monolayer transferred onto a SiO$_2$/Si substrate for increasing density of excitons created per pulse. The fits were obtained assuming 11.5% absorption of the optical excitation at 405 nm[57] and accounting for the optical losses in the measurement setup. The inset shows the

PL's initial decay emphasizing the lack of excitation density-induced decay saturation. (b) Auger constant estimated by fitting the data in (a) with Eq.(2). The y-axis represents the product of the initial exciton density denoted by $n(0)$ and the Auger constant $\gamma_A$. (c) Integrated PL of the h-BN encapsulated WSe$_2$ monolayer for increasing density of excitons created per pulse. The Auger constant was estimated by fitting the PL intensity with the relation $I_{PL} \propto \ln[1 + n(0)\gamma_A\tau](\gamma_A\tau)^{-1}$ [3]. The solid lines indicate the fits, and the markers the experimental data.

The estimated values for Auger constant and exciton lifetime from our TRPL measurements are 0.02 cm$^2$s$^{-1}$ and 0.23 ns respectively as shown in Figure 2(b-c). Based on these values, the contribution of Auger broadening to the MSD should be minimal for excitation densities lower than $2\times 10^{11}$ cm$^{-2}$ (i.e. $(\gamma_A\tau)^{-1}$). To confirm this prediction, the Auger constant obtained in Figure 2 was used to estimate the degree of Auger broadening in our measurements at various excitation densities and found very little contribution to the total MSD for excitation densities below $4\times 10^{11}$ cm$^{-2}$ (see supplementary material). This observation is consistent with other reports where exciton-exciton Auger scattering has been shown to be drastically reduced in h-BN encapsulated TMD monolayers[45]. More importantly, Auger broadening cannot explain the saturation observed in Figure 1 at excitation densities above $2\times 10^{11}$cm$^{-2}$ as the integrated PL intensity continues to rise for excitation densities above this value as shown in Figure 2(c). Furthermore, the Auger coefficient extracted from the integrated PL intensity using the relation $I_{PL} \propto \ln[1 + n(0)\gamma_A\tau](\gamma_A\tau)^{-1}$ from reference[3] is consistent with the Auger coefficient obtained from our TRPL measurements. Therefore, the fast rise of the MSD at early times is caused predominantly by the fast motion of *hot* excitons whose average temperature is higher than the lattice temperature[48,50,51,53,58]. Consequently, the progression from fast to slow evolution of the MSD for a given excitation density can be explained as the transition from a *hot* gas of excitons, which move very fast due to their high kinetic energy, to a *cold* gas of excitons that move more slowly after having relaxed their excess kinetic energy.

The distribution of kinetic energy immediately after the formation of *hot* excitons is very far from equilibrium. This distribution evolves towards quasi-equilibrium described by an effective temperature via scattering with other *hot* excitons and with phonons[29]. After a thermalized distribution is achieved, and before recombination occurs, *hot* excitons typically lose their excess kinetic energy to the lattice via phonon scattering[29]. In the high excitation density regime, however, the cooling of *hot* excitons by exciton-phonon scattering is counteracted by the heating of the exciton gas via exciton-exciton Auger scattering[46,47]. During this Auger process, an exciton recombines non-radiatively and transfers its energy to a nearby exciton which ionizes into *hot* carriers with excess energy of the order of the bandgap energy. These *hot* carriers lose their excess energy again via carrier-phonon and carrier-carrier scattering and eventually form new excitons with kinetic energies that may be higher than the originally-formed excitons' kinetic energy. These newly-formed *hotter* excitons then scatter with the remaining colder excitons thereby increasing the overall gas temperature[46]. A schematic illustration of the exciton formation and Auger heating processes, as well as a timescale of the exciton formation, relaxation and transport regimes are shown in Figure 3.

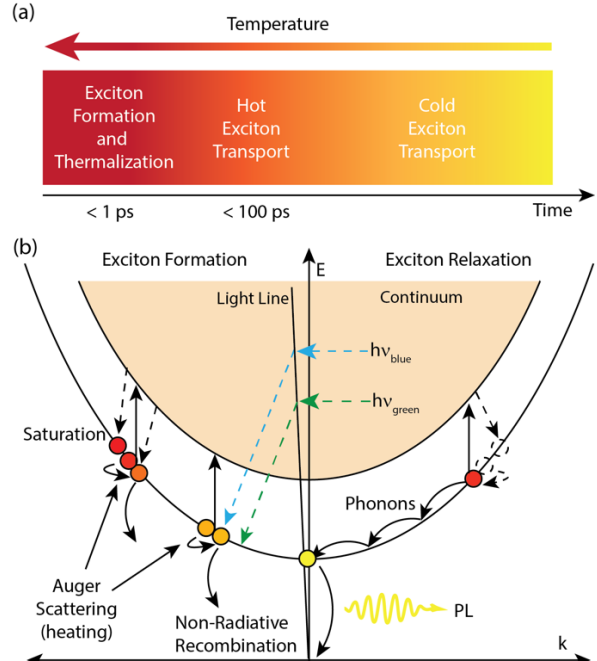

Figure 3: Schematic diagram illustrating (a) the timescales of exciton formation, *hot* and *cold* exciton transport regimes as well as the evolution of the temperature of the exciton gas, and (b) exciton dispersion relation showing the exciton formation and relaxation processes. The saturation refers to the rate balance between exciton-exciton Auger scattering and exciton-phonon scattering.

Auger heating depends strongly on the density of the exciton population, which implies that it is most likely to occur early in the relaxation process when the density of the exciton gas is at its highest. While *hot* excitons relax their kinetic energy, they can also propagate and while doing so their speed should decrease. After the exciton density has decayed enough for Auger scattering to become less relevant, and the excess kinetic energy of the exciton gas has been relaxed to the lattice via exciton-phonon scattering, the remaining *cold* excitons diffuse at a constant rate. This trend is consistent with our observations of the nonlinear evolution of the MSD of h-BN encapsulated WSe$_2$ monolayers shown in Figure 1.

The evolution of the MSD of a gas of excitons created by an optical pulse with a Gaussian spatial profile can be modeled as $\langle\Delta\sigma^2(t)\rangle = 2tD(t)$, where the time-varying diffusivity $D(t)$ depends on the instantaneous effective temperature of the exciton gas $T(t)$[59]. Assuming that the diffusivities and mobilities of the bound electron and hole making up an exciton are identical, the instantaneous exciton diffusivity can be expressed explicitly in terms of the exciton temperature as $D(t) = \mu_q k_B T(t)/q$ where $\mu_q$ represents the mobility of a carrier with charge $q$, and $k_B$ is the Boltzmann constant. Since the energy of the exciton gas is relaxed to the lattice via phonon scattering, one can model the evolution of the exciton temperature as

$$T(t) = T_0 + T^* e^{-t/\tau^*} \quad (3)$$

where $T_0$, $T^*$ and $\tau^*$ represent the exciton gas steady-state temperature, the initial excess temperature, and the kinetic energy relaxation time constant respectively. If the exciton gas completely relaxes its excess kinetic energy before it recombines, the steady-state temperature should equal the lattice temperature, and the kinetic energy relaxation time constant is approximately equal to the lifetime. With these assumptions, the MSD of the exciton gas in a h-BN encapsulated WSe$_2$ monolayer was fitted and the resulting excess temperatures are shown in Figure 4. Following the same trend as the slope of the MSD, the fitted excess temperature $T^*$ appears to increase as the excitation density increases, and it also saturates at elevated excitation densities. This saturation is expected to occur due to the balancing effects of Auger heating and phonon cooling of the exciton gas discussed earlier. It is well known that the carrier or exciton-phonon scattering rate increases as the energy of the carrier or exciton increases[60]. This implies that exciton-phonon scattering is more likely to dominate the energy transfer mechanism that determines the temperature of the exciton gas at high excitation densities. Therefore, the observed saturation of the initial slope of the MSD and the extracted initial excess gas temperatures (Figure 4), is consistent with the balance between exciton-phonon scattering, which removes energy from the exciton gas, and Auger heating, which increases the energy of the exciton gas.

To further confirm our conclusion, we performed an experiment on a different h-BN encapsulated WSe$_2$ monolayer with lower excitation photon energy. If the nonlinear evolution of the MSD is indeed caused by the relaxation of kinetic energy of a *hot* exciton gas and not Auger broadening, then for lower excitation energies, the saturation of the initially-fast MSD slope and initial excess exciton temperature should occur at higher excitation densities. This is expected because the higher the excitation energy, the higher the kinetic energy of the initially formed *hot* exciton gas, and the less energy that will be required via Auger heating to reach the saturation temperature. This trend is precisely evidenced in Figure 4 where the excess temperature of the initially formed exciton gas is shown for two different excitation energies, 3.1 eV (405 nm) and 2.4 eV (520 nm), and increasing excitation densities.

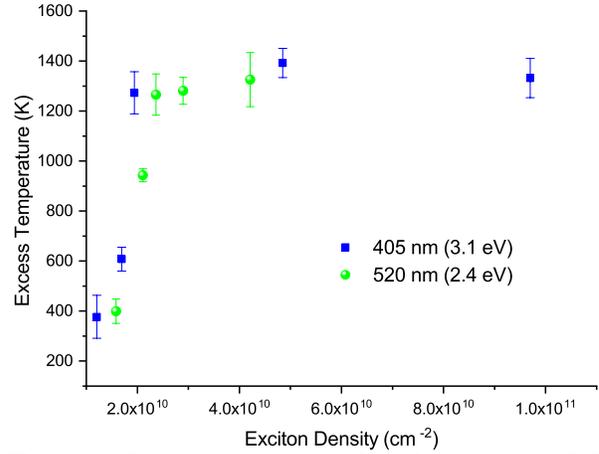

Figure 4: Excess temperature of excitons in a h-BN encapsulated WSe$_2$ monolayer created by two lasers with different photon energies and increasing excitation densities obtained by fitting the MSD with the model $\langle\Delta\sigma^2(t)\rangle = 2tD(t) = 2t\mu_q k_B/qT(t)$ where the time-dependent temperature is given by Eq.(3). The saturation of the excess temperature of the exciton gas occurs at a higher excitation density for the optical excitation with lower photon energy (2.4 eV) than for the optical excitation with higher photon energy (3.1 eV). The error bars correspond to the 95% confidence interval of the fits. These data correspond to a sample that is different from that shown in Figure 1.

As expected, the excitation density required to reach the saturation excess temperature with the lower photon energy excitation (520 nm) is higher than the excitation density required to reach the same saturation temperature with the higher photon energy excitation (405 nm). This correlation between excitation density and photon energy is convincing evidence that the initial fast motion of the exciton gas stems primarily from the motion of *hot* excitons created by non-resonant excitation rather than Auger broadening.

In summary, through a systematic investigation via temporally and spatially resolved PL measurements at room temperature, we have determined the origin of the apparently anomalous diffusive motion of excitons in h-BN encapsulated WSe$_2$ monolayers. Specifically, we have shown that the nonlinear evolution of the MSD of an exciton gas in a h-BN encapsulated WSe$_2$ monolayer created by a high-density, non-resonant optical excitation is dominated by *hot* exciton transport. We observed a correlation between the initial slope of the MSD and the excitation density that saturates at elevated excitation densities. This saturation is consistent with the balancing effects of Auger heating and phonon cooling that respectively increase and decrease the temperature of the exciton gas. We confirmed that the excitation density required to reach the saturation temperature depends on the excitation's photon energy. That is, a higher photon energy excitation requires a lower excitation density to reach the saturation temperature. These results offer new insight into the exciton transport dynamics in TMDs that will aid in the design of excitonic devices that exploit the regimes of *hot* and *cold* exciton transport.

See supplementary material for details of the fabrication processes, and data analysis.

This work was supported through the AFOSR grant No. 16RT1256 and startup grant from the University of Michigan. D.C L would like to acknowledge support from the University of Michigan's Rackham Merit Fellowship and the National Science Foundation Graduate Research Fellowship Program under Grant No. DGE 1256260. Any opinions, findings, and conclusions or recommendations expressed in this material are those of the author(s) and do not necessarily reflect the views of the National Science Foundation. All authors contributed to and approved the final version of the manuscript.